\documentclass[conference]{IEEEtran}
\IEEEoverridecommandlockouts

\usepackage[utf8]{inputenc}
\usepackage{amsmath, amssymb, amsfonts}
\usepackage{algorithmic}
\usepackage{algorithm}
\usepackage{graphicx}
\usepackage[caption=false,font=footnotesize]{subfig}
\usepackage{textcomp}
\usepackage{xcolor}
\usepackage{cite}
\usepackage{booktabs}
\usepackage{multirow}
\usepackage{balance}
\usepackage{hyperref}

\usepackage{array}         
\usepackage{tabularx}      
\usepackage[table]{xcolor}        
\usepackage{bm}
\usepackage{mathtools}
\usepackage{flushend}
\newcommand{\THz}{\text{THz}}
\newcommand{\GHz}{\text{GHz}}

\newcolumntype{L}[1]{>{\raggedright\arraybackslash}p{#1}}
\newcolumntype{C}[1]{>{\centering\arraybackslash}p{#1}}
\newcolumntype{R}[1]{>{\raggedleft\arraybackslash}p{#1}}

\definecolor{rowgray}{gray}{0.93}
\definecolor{hdrblue}{RGB}{45,74,122}
\definecolor{secblue}{RGB}{45,74,122}

\newcommand{\dB}{\,\text{dB}}
\newcommand{\dBm}{\,\text{dBm}}
\newcommand{\dBi}{\,\text{dBi}}
\newcommand{\Npm}{\,\text{Np/m}}

\newcommand{\Nact}{N_{\mathrm{act}}}
\newcommand{\Pt}{P_t}

\newcommand{\Tsys}{T_{\mathrm{sys}}}

\newcommand{\neff}{n_{\mathrm{eff}}}

\begin{document}

\title{Pinching-Antenna-Assisted Terahertz Communications: Modeling and Benchmarking\\
\thanks{This work was supported by the German Federal Ministry of Education and Research (BMBF) through \emph{Open6GHub+} (Grant no.  \emph{16KIS2402K}) project.}
}
\author{\IEEEauthorblockN{Wei Jiang}
\IEEEauthorblockA{\textit{Intelligent Networking Research Group}\\
\textit{German Research Center for Artificial Intelligence (DFKI)} \\
Kaiserslautern, 67663 Germany}
\and
\IEEEauthorblockN{Hans D. Schotten}
\IEEEauthorblockA{\textit{Department of
Electrical and Computer Engineering}\\
\textit{University of Kaiserslautern (RPTU)} \\
Kaiserslautern, 67663 Germany}
}

\maketitle

\begin{abstract}
Pinching antenna systems (PASS) employing dielectric waveguides have recently emerged as a promising flexible antenna architecture for high-frequency wireless communications. While prior work has focused primarily on millimeter-wave regimes, extending PASS to the terahertz (THz) band introduces distinct electromagnetic phenomena that invalidate conventional modeling assumptions. This paper develops the first analytical framework for THz-PASS that integrates in-waveguide propagation attenuation, evanescent coupling via coupled-mode theory, and THz-specific free-space effects including molecular absorption and its re-radiation noise. Using this model, we benchmark THz-PASS against conventional phased arrays under identical propagation scenarios. Our comparative evaluation reveals that THz-PASS achieves effective gains in spectral efficiency through proximity exploitation, making it particularly well-suited for confined and linear deployment topologies. 
\end{abstract}


\section{Introduction}

Terahertz (THz) frequencies offer enormous bandwidth for next-generation wireless networks \cite{Ref_jiang2024terahertz}. However, THz radiation faces fundamental propagation challenges: free-space path loss scales quadratically with frequency, atmospheric absorption introduces severe attenuation, and even common materials cause serious blockage \cite{akyildiz2022terahertz}. The conventional approach to overcome these impairments relies on massive antenna arrays at the transceiver, employing hundreds to thousands of fixed-position elements. Through digital, analog, or hybrid beamforming \cite{molisch2017hybrid}, these arrays generate pencil beams that concentrate energy toward users, providing high directional gain and—when equipped with multiple RF chains—spatial multiplexing capability \cite{jiang2023thz}. While effective, this paradigm incurs substantial hardware complexity, cost, and power consumption that scale with array size, motivating the exploration of fundamentally novel antenna architectures for THz systems.

Recently, pinching antenna systems (PASS) have garnered significant attention as a flexible antenna architecture for high-frequency wireless communications \cite{liu2025magazine}. Originally prototyped by NTT DOCOMO in 2021 \cite{fukuda2021pinching}, PASS employs dielectric waveguides as the transmission medium, while small dielectric particles, known as pinching antennas (PAs), can be dynamically attached at arbitrary points along the waveguide—ideally positioned in close proximity to each user. Each pinched element acts as a localized radiator, coupling a portion of the guided wave into free space via evanescent field leakage. By drastically reducing the  free-space propagation distance and enabling dynamic blockage circumvention, PASS establishes robust, short-range line-of-sight (LoS) links \cite{ding2025flexible, liu2025pinching}. These attributes position PASS as a compelling candidate to address the severe propagation challenges inherent in THz-band communications.

While prior research on PASS has predominantly focused on millimeter-wave frequencies, extending this architecture to the THz band introduces fundamentally different electromagnetic phenomena that remain unexplored in existing PASS works \cite{ouyang2026multicast}. To assess the viability of THz-PASS and provide quantitative design insights, this paper presents the first systematic comparative analysis of conventional phased arrays and pinching-antenna systems for THz communications. Our main works include:

\begin{itemize}
    \item \textit{Unified analytical framework:} We develop an analytical framework for THz-PASS that integrates dielectric absorption, surface-roughness-induced scattering, evanescent coupling via coupled-mode theory, and THz-specific free-space effects including molecular absorption and its re-radiation noise. Moreover, we derive a sequential depletion model for power distribution along the waveguide that captures the in-waveguide attenuation. 
    \item \textit{Comparative performance benchmarking:} Through extensive Monte Carlo simulations under identical propagation conditions, we quantify the spectral efficiency trade-offs between THz-PASS and conventional phased arrays, identifying crossover points where each option excels.
\end{itemize}

The remainder of this paper is organized as follows.
Section II presents unified system models for both
architectures. Section III develops the THz-PASS propagation characterization.
Section IV formulates achievable performance. Section V provides numerical results. Section VI concludes the paper.

\begin{figure}[!t]
\centering 
\includegraphics[width=0.98\linewidth]{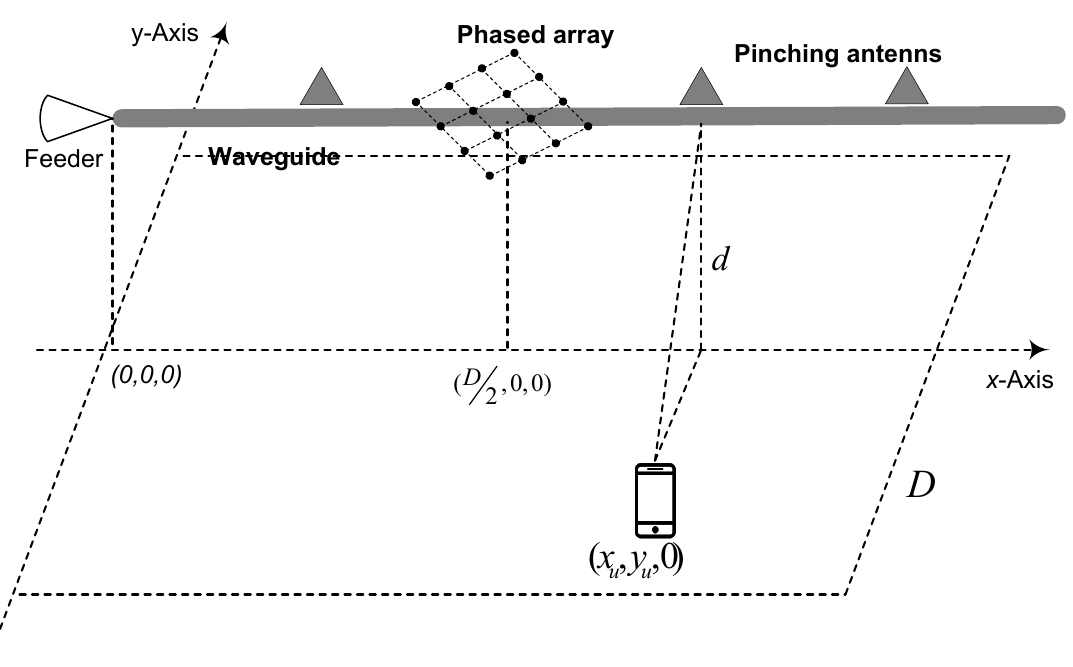}
\caption{Geometric configuration for PASS-THz systems. The waveguide is placed parallel to the $x$-axis at height $d$; pinching antennas are movable along the waveguide. In a conventional THz system, the UPA is fixed at the center.}
\label{fig:geometry}
\end{figure}

\section{System Model}

We consider a downlink THz communication scenario where a single-antenna user equipment (UE) is served by a base station (BS). A pinching-antenna system employing a reconfigurable dielectric waveguide is modeled and compared with a conventional uniform planar array (UPA) with fixed-position elements. Both systems operate at the same carrier frequency $f_c$ in the sub-THz band (100--300~\GHz), or the THz band (300\GHz --3~\THz), and are evaluated under identical propagation conditions. 
As illustrated in \figurename~\ref{fig:geometry}, a three-dimensional Cartesian coordinate system is adopted. The UE is located on the ground plane ($z=0$) at coordinates $\bm{\psi}_{\mathrm{u}} = (x_u, y_u, 0)$. The UE is uniformly distributed within the coverage area.

The PASS employs a single low-loss dielectric waveguide of length $L$ placed parallel to the $x$-axis at height $d$ and lateral position $y=0$. A single feed point is located at $\bm{\psi}_0 = (0,0,d)$. Along the waveguide, each PA is a small dielectric particle that can be dynamically attached to the waveguide surface. Following the physics-based hardware model established in recent literature \cite{wang2025modeling}, the PA couples a fraction of the guided wave into free space via evanescent field leakage, with the coupling mechanism modeled using coupled-mode theory. The position of the $n$-th activated PA is denoted $\bm{\psi}_n^{\mathrm{PA}} = (x_n, 0, d)$, $n = 1,\ldots,N_{\mathrm{act}}$, where $N_{\mathrm{act}} $ is the number of activated PAs. 
The UPA is installed at a fixed location. Its geometric center is placed at $\bm{\psi}_a = (L/2,0,d)$, directly above the center of the user distribution area. The array consists of $M = M_x \times M_y$ elements arranged on a rectangular grid with inter-element spacing $d_x = d_y = \lambda_0/2$, where $\lambda_0 = c/f_c$ is the free-space wavelength. The position of the $(i,j)$-th element ($i = 1,\ldots,M_x$, $j = 1,\ldots,M_y$) is
\begin{equation}
\bm{\psi}_{i,j}^{\mathrm{UPA}} = \left( \frac{L}{2}+\left(i-\frac{M_x+1}{2}\right)d_x,\; \left(j-\frac{M_y+1}{2}\right)d_y,\; d \right).
\label{eq:upa_position}
\end{equation}

\section{In-Waveguide and Free-Space Propagation Modeling of THz-PASS}
We first establish a unified propagation model that captures both guided-wave transmission inside the dielectric waveguide and subsequent free-space radiation toward the user.

\subsubsection{In-Waveguide Propagation}
The guided wave propagates along the $+x$ direction. The phase constant is $\beta = 2\pi / \lambda_g$, where $\lambda_g = \lambda_0 / n_{\mathrm{eff}}$ is the guided wavelength and $n_{\mathrm{eff}}$ is the effective refractive index of the dielectric waveguide\footnote{While the dielectric waveguide inherently exhibits group velocity dispersion at THz frequencies, this analytical framework assumes a narrowband signal where the coherence bandwidth is significantly larger than the signal bandwidth. Thus, the phase constant $\beta(f)$ is approximated as linear over the frequency range of interest, neglecting pulse broadening for the baseline performance analysis.}. For all-dielectric waveguides used in PASS, conductor losses are absent. The attenuation constant $\alpha$ accounts for dielectric absorption and surface-roughness-induced scattering — the latter being particularly pronounced at THz frequencies due to the short wavelength and high contrast between core and cladding materials. The signal at a distance $x$ from the feed is
\begin{equation}
s(x) = s_0 \, e^{-(\alpha + j\beta) x},
\label{eq:waveguide_field}
\end{equation} where $s_0$ is the information symbol with $\mathbb{E}[|s_0|^2]=1$.

\subsubsection{Pinching Antenna Coupling}
A PA placed at $x_n$ couples a portion of the guided power into free space. Following the physics-based hardware model in \cite{wang2025modeling, ding2025flexible}, the pinching antenna is modeled as an open-ended directional coupler, and the electromagnetic field behavior is analyzed using coupled-mode theory. For analytical tractability, two power allocation strategies for multiple PAs on a shared waveguide are considered \cite{xu2025pinching, liu2025pinching}:

\begin{itemize}
    \item \textit{Equal Power Model:} The coupling gap/length of each PA is individually tuned  such that every activated antenna radiates an identical amount of power ($P_1 = P_2 = \dots = P_{N_{\mathrm{act}}}$), compensating for the progressive depletion of the guided wave along the waveguide.
    \item \textit{Proportional Power Model:} All PAs employ an identical coupling gap/length, meaning each PA radiates a fixed fraction of the local waveguide power available at its specific position. This model reflects the physical reality of a simple implementation where all PAs are identical.
\end{itemize}

In a series-fed PASS, the power reaching the $n$-th PA is reduced by two factors: waveguide attenuation over the propagation distance $x_n$\footnote{The existing power model for PASS (e.g., \cite{liu2025pinching}) neglects waveguide loss, an assumption that is valid at low and millimeter-wave frequencies but becomes inaccurate at THz frequencies.}, and power extraction by all $n-1$ preceding PAs. Under the proportional power model (assuming PAs are ordered $x_1 < x_2 < \dots < x_{N_{\mathrm{act}}}$), the power radiated by the $n$-th element follows a sequential depletion process:
\begin{equation}
P_n = P_t \cdot \kappa^2 \cdot e^{-2\alpha x_n} \cdot (1 - \kappa^2)^{n-1},
\label{eq:sequential_power_fixed}
\end{equation}
where $P_t$ is the transmit power at the feed ($x=0$) and $\kappa^2$ is the uniform coupling efficiency. 
This formulation captures the essential physics: each PA radiates power proportional to the local waveguide power at its position, with the local power having been reduced by both attenuation along the waveguide and power extraction by all previous PAs. Crucially, the depletion factor $(1-\kappa^2)^{n-1}$ depends only on the number of preceding couplers, not on the distances between them — once power is coupled out at a given location, it is permanently removed from the waveguide regardless of subsequent propagation distances.
For the purpose of this comparative study, we focus on the scenario where PAs are clustered near the optimal point $x_u$. When the cluster span is small such that $|x_n - x_{n-1}| \ll 1/\alpha$ and $|x_n - x_u| \ll 1/\alpha$, the attenuation factor $e^{-2\alpha x_n} \approx e^{-2\alpha x_u}$ is approximately constant across the cluster. 

Furthermore, for modest $N_{\mathrm{act}}$ and typical coupling coefficients ($\kappa^2 \ll 1$), which satisfies $N_{\mathrm{act}} \cdot \kappa^2 \leq 1$, the depletion factor $(1-\kappa^2)^{n-1} \approx 1$ for all PAs in the cluster. Under these conditions, the power radiated by each PA simplifies to
\begin{equation}
P_n \approx \frac{P_t}{N_{\mathrm{act}}} \cdot \kappa^2 \cdot e^{-2\alpha x_u},
\label{eq:proportional_power_approx}
\end{equation}
where we have implicitly assumed that the total power available at the cluster is approximately $P_t e^{-2\alpha x_u}$ and is equally shared among the $N_{\mathrm{act}}$ PAs due to their proximity and identical coupling. This approximation, valid for the clustered deployment, captures the fundamental THz trade-off: the advantage of reducing free-space path loss via proximity ($x_u$) versus the accumulation of waveguide-induced attenuation.

\subsubsection{Free-Space Radiation}
Each PA radiates as a point source with directivity \cite{ding2025flexible}
\begin{equation}
G_{\mathrm{PA}} = \frac{4\pi \epsilon_{r,\mathrm{PA}} V_{\mathrm{PA}}}{\lambda_0^3},
\label{eq:pa_directivity}
\end{equation}
where $\epsilon_{r,\mathrm{PA}}$ is the relative permittivity of the PA material and $V_{\mathrm{PA}}$ is its volume. The free-space channel between the $n$-th PA and the UE is 
\begin{equation}
h_n^{\mathrm{fs}} = \sqrt{ G_{\mathrm{PA}} G_r } \, \frac{c}{4\pi f_c d_n} \, e^{-j\frac{2\pi}{\lambda_0} d_n} \cdot \sqrt{ \tau_{\mathrm{abs}}(f_c, d_n) },
\label{eq:pass_fs_channel}
\end{equation}
where $d_n = \|\bm{\psi}_n^{\mathrm{PA}} - \bm{\psi}_{\mathrm{u}}\|$ denotes the propagation distance between the $n^{th}$ PA and the user. THz channel modeling must incorporate phenomena negligible at lower frequencies \cite{bhardwaj2024generalized, jornet2011channel}, especially molecular absorption, which is frequency- and distance-dependent attenuation:
\begin{equation}
\tau_{\text{abs}}(f,d) = e^{-K(f)d}
\label{eq:absorption}
\end{equation}
where $K(f) = \sum_i n_i \sigma_i(f)$, $n_i$ is the number density of gas $i$ (in molecules/m$^3$) and $\sigma_i(f)$ is its frequency-dependent absorption cross-section (in m$^2$/molecule). Following \cite{bhardwaj2024generalized}, we model the absorption coefficient as a Gamma-distributed random variable:
\begin{equation}
K(f) \sim \Gamma(\epsilon_K, \eta_K)
\label{eq:random_absorption}
\end{equation}
to account for atmospheric variability.

\subsubsection{Noise Model with Molecular Absorption}
In addition to thermal noise, THz channels experience molecular absorption noise due to re-radiation by excited molecules. Following the established model \cite{jornet2011channel}, the total one-sided power spectral density is
\begin{equation}
N_{\text{total}}(f,d) = k_B T_{\text{sys}} + k_B T_0 \left(1 - e^{-K(f)d}\right),
\label{eq:noise_psd}
\end{equation}
where $T_{\text{sys}}$ is the system noise temperature and $T_0 = 290$ K is the reference ambient temperature. The noise power therefore be computed as $\varsigma^2=\int_B N_{\text{total}}(f,d) df$, where $B$ is the system bandwidth. This effect is particularly significant near molecular absorption peaks and narrows the usable transmission windows.

\section{Comparative Performance Analysis}
This section comparatively analyzes the achievable spectral efficiency of THz-PASS and phased arrays.

\subsection{THz-PASS Systems}
Because all PAs are fed from the same waveguide, the transmitted signals are phase-coherent but inherit the propagation phase shift accumulated within the waveguide. Let $\theta_n = \beta x_n$ denote the phase delay from the feed to the $n$-th PA (with $x_{\mathrm{feed}}=0$). Using the clustered deployment approximation from \eqref{eq:proportional_power_approx}, the signal radiated by the $n$-th PA is $\sqrt{P_n} \, e^{-j\theta_n} s_0$ with $P_n$ given by \eqref{eq:proportional_power_approx}. The UE receives the superposition
\begin{equation}
y = \sqrt{P_t} \sum_{n=1}^{N_{\mathrm{act}}} \sqrt{ \frac{\kappa^2 e^{-2\alpha x_u}}{N_{\mathrm{act}}} } \, h_n^{\mathrm{fs}} \, e^{-j\theta_n} \, s_0 + n_{\mathrm{total}},
\label{eq:pass_rx}
\end{equation}
where $n_{\mathrm{total}}$ represents noise with PSD given by \eqref{eq:noise_psd}. The effective channel coefficient is defined as
\begin{equation} \label{GS_effCHgain}
h_{\mathrm{PASS}} =  \frac{\kappa e^{-\alpha x_u}}{\sqrt{ N_{\mathrm{act}}} } \sum_{n=1}^{N_{\mathrm{act}}} h_n^{\mathrm{fs}} \, e^{-j\theta_n}.
\end{equation}
The received SNR is $\gamma_{\mathrm{PASS}} = P_t |h_{\mathrm{PASS}}|^2 / \varsigma^2$.

A unique advantage of PASS is the ability to position PAs to simultaneously minimize free-space propagation distances and align the phases for coherent combining. For the single-user scenario, the optimal strategy involves a two-stage process \cite{wang2025modeling}:

\begin{enumerate}
    \item \textit{Large-scale placement:} PAs are clustered around the location on the waveguide closest to the UE, i.e., $x_n = x_u + \Delta_n$, where $\Delta_n$ are small offsets from the user's projection point. This clustering minimizes the free-space path loss by ensuring all PAs are in close proximity to the UE.
    \item \textit{Phase alignment:} With PAs clustered near $x_u$, their positions are finely adjusted to ensure coherent combining at the UE. The total phase for the $n$-th PA is the sum of the waveguide propagation phase $\theta_n = \beta x_n$ and the free-space propagation phase $(2\pi/\lambda_0) d_n$. Assuming narrowband operation such that $\beta$ is approximately constant over the signal bandwidth, the phase-matching condition is
    \begin{equation}
    \frac{2\pi}{\lambda_0} d_n + \beta x_n = 2k\pi + \phi_0, \quad k \in \mathbb{Z},
    \label{eq:phase_matching}
    \end{equation}
    where $\phi_0$ is a common phase reference. For wideband systems, waveguide dispersion would require more sophisticated frequency-dependent phase compensation.
\end{enumerate}

The distance from a PA at $x_n$ to the UE is
\begin{equation}
d_n = \sqrt{(x_n - x_u)^2 + y_u^2 + d^2}.
\label{eq:pa_distance}
\end{equation}
When PAs are placed directly above the user (i.e., $x_n = x_u$), this reduces to the minimum possible distance
\begin{equation}
d_{\min} = \sqrt{y_u^2 + d^2}.
\label{eq:dmin}
\end{equation}
With perfect phase alignment, the effective channel gain in \eqref{GS_effCHgain} becomes
\begin{align} \nonumber
|h_{\mathrm{PASS}}|^2 \approx & \frac{\kappa^2 e^{-2\alpha x_u}}{N_{\mathrm{act}}} G_{\mathrm{PA}} G_r \left( \frac{c}{4\pi f_c d_{\min}} \right)^2 \tau_{\mathrm{abs}}(f_c, d_{\min}) \\
& \cdot \left| \sum_{n=1}^{N_{\mathrm{act}}} e^{-j(\frac{2\pi}{\lambda_0}(d_n - d_{\min}) + \beta \Delta_n)} \right|^2.
\label{eq:pass_gain_approx}
\end{align}

For moderate $N_{\mathrm{act}}$ and small offsets $\Delta_n$, the phase terms in the summation approach zero, yielding coherent combining that approaches $N_{\mathrm{act}}^2$. However, it is important to note that mutual coupling between closely spaced PAs prevents the array gain from scaling linearly with $N_{\mathrm{act}}$ indefinitely. Recent research demonstrates the existence of an optimal number of antennas $N^*$ that maximizes the array gain, beyond which mutual coupling degrades performance \cite{ouyang2025ieee}. Accounting for this effect, the gain can be expressed as
\begin{align} \label{eq:pass_gain_final}
&|h_{\mathrm{PASS}}|^2 \approx \\ \nonumber & \eta(N_{\mathrm{act}}) \, \kappa^2 e^{-2\alpha x_u} G_{\mathrm{PA}} G_r \left( \frac{c}{4\pi f_c d_{\min}} \right)^2 \tau_{\mathrm{abs}}(f_c, d_{\min}),
\end{align}
where $\eta(N_{\mathrm{act}})$ is an effective array gain factor that accounts for coherent combining gains tempered by mutual coupling losses. For small $N_{\mathrm{act}}$ (e.g., $N_{\mathrm{act}} \le 8$), $\eta(N_{\mathrm{act}}) \approx N_{\mathrm{act}}$ is a reasonable approximation, but for larger arrays, the sub-linear scaling must be considered. Thus, PASS achieves a power gain through coherent combining, analogous to the array gain of conventional systems, but with the crucial distinction that the gain is realized by drastically reducing the effective propagation distance from $d = \|\bm{\psi}_0 - \bm{\psi}_{\mathrm{u}}\|$ to $d_{\min} \le d$, in addition to the multiplicity of elements.

\subsection{Benchmark System: Conventional Phased Arrays}
The UPA employs baseband beamforming. Let $\mathbf{w} \in \mathbb{C}^M$ denote the unit-norm beamforming vector ($\|\mathbf{w}\|^2 = 1$). The transmitted signal is $\mathbf{x} = \sqrt{P_t} \mathbf{w} s_0$, and the received signal at the UE is
\begin{equation}
y = \sqrt{P_t} \, \mathbf{h}^H \mathbf{w} \, s_0 + n_{\mathrm{total}},
\label{eq:upa_rx}
\end{equation}
where $n_{\mathrm{total}}$ represents noise with PSD given by \eqref{eq:noise_psd} and $\mathbf{h} \in \mathbb{C}^M$ is the channel vector. At THz frequencies, the channel is dominated by the LoS path. Measurement campaigns have shown that non-LoS components are typically 20-30 dB weaker at sub-THz frequencies (140 GHz) and can be 30-50 dB or more below the LoS path at higher THz frequencies ($>$300 GHz) due to increased material absorption and reflection loss \cite{rappaport2019wireless, vegni2016chirality}. For the frequency range considered in this work, we therefore neglect NLoS components as a valid approximation.

The channel coefficient between the $(i,j)$-th UPA element and the UE is
\begin{equation}
h_{i,j} = \sqrt{ G_t G_r } \, \frac{c}{4\pi f_c d_{i,j}} \, e^{-j\frac{2\pi}{\lambda_0} d_{i,j}} \cdot \sqrt{ \tau_{\mathrm{abs}}(f_c, d_{i,j}) },
\label{eq:upa_channel_element}
\end{equation}
where $d_{i,j} = \|\bm{\psi}_{i,j}^{\mathrm{UPA}} - \bm{\psi}_{\mathrm{u}}\|$ is the Euclidean distance, $G_t$ and $G_r$ are the antenna gains of each BS element and the UE, respectively. The channel vector $\mathbf{h}$ is obtained by stacking $h_{i,j}$. 
Assuming perfect channel information, the optimal beamforming vector to maximize the received SNR is maximum-ratio transmission (MRT) $\mathbf{w}_{\mathrm{opt}} = \frac{\mathbf{h}}{\|\mathbf{h}\|}$. 
The resulting SNR is
\begin{equation}
\gamma_{\mathrm{UPA}} = \frac{P_t}{\varsigma^2} \|\mathbf{h}\|^2.
\label{eq:upa_snr}
\end{equation}

\subsection{Unified SNR Expression}
To facilitate a direct comparison, we cast the SNR for both architectures into a common form:
\begin{equation}
\gamma = \frac{P_t}{\varsigma^2} \cdot G_{\mathrm{eff}} \cdot \left( \frac{c}{4\pi f_c d_{\mathrm{eff}}} \right)^2 \cdot \mathbb{E}[\tau_{\mathrm{abs}}(f_c, d_{\mathrm{eff}})] \cdot \xi,
\label{eq:snr_unified}
\end{equation}
where
\begin{itemize}
    \item $G_{\mathrm{eff}}$ is the effective antenna gain (including beamforming gain, element gain, coupling efficiency, and the UE receiver gain);
    \item $d_{\mathrm{eff}}$ is the effective propagation distance;
    \item $\mathbb{E}[\tau_{\mathrm{abs}}(f_c, d_{\mathrm{eff}})]$ is the expected molecular absorption loss given the statistical nature of $K(f)$;
    \item $\xi$ captures additional losses (waveguide attenuation).
\end{itemize}

For the UPA with maximum-ratio transmission, the channel norm is $\|\mathbf{h}\|^2 = \sum_{i,j} |h_{i,j}|^2$. Using \eqref{eq:upa_channel_element} and assuming all $d_{i,j} \approx d_{\mathrm{eff}} = \|\bm{\psi}_0 - \bm{\psi}_{\mathrm{u}}\|$, we obtain
\[
\|\mathbf{h}\|^2 \approx M G_t G_r \left( \frac{c}{4\pi f_c d_{\mathrm{eff}}} \right)^2 \tau_{\mathrm{abs}}(f_c, d_{\mathrm{eff}}).
\]
Thus, for the UPA,
\[
G_{\mathrm{eff}} = M G_t G_r, \qquad d_{\mathrm{eff}} = \|\bm{\psi}_0 - \bm{\psi}_{\mathrm{u}}\|, \qquad \xi = 1.
\]
With phase alignment and clustering near the user's projection, from \eqref{eq:pass_gain_final}, we know that, for PASS,
\[
G_{\mathrm{eff}} = \eta(N_{\mathrm{act}}) \kappa^2 G_{\mathrm{PA}} G_r, \qquad d_{\mathrm{eff}} = d_{\min} = \sqrt{y_u^2 + d^2},
\] and 
\[ \qquad \xi = e^{-2\alpha x_u}. \]
This unified representation highlights the fundamental trade-off: conventional arrays achieve gain through spatial superposition of many fixed elements, whereas PASS achieves gain through a combination of proximity (reduced $d_{\mathrm{eff}}$) and coherent combining of few reconfigurable elements.

\section{Performance Evaluation}
\label{sec:results}

We present a numerical evaluation of the THz-PASS and UPA
architectures under realistic THz propagation conditions. We simulate a downlink indoor  environment at a carrier frequency of $f_c = 300\GHz$. By default, the room measures $20\,\text{m} \times 20\,\text{m}$ with a ceiling height of $3\,\text{m}$. All
results are obtained through Monte Carlo simulation with $2000$
independent drops of uniformly distributed random UE positions on the floor plane and
random atmospheric absorption realizations drawn from the Gamma-distributed model in~\eqref{eq:random_absorption}. A dielectric waveguide is ceiling-mounted ($d=3\,\text{m}$) parallel to the $x$-axis. The UPA is co-located at the same height with its geometric center
directly above the room center. Table~\ref{tab:sim_params} summarizes the complete simulation parameters, grouped by category.

\begin{table}[!t]
  \centering
  \caption{Simulation Parameters}
  \label{tab:sim_params}
  \renewcommand{\arraystretch}{1.20}
  \setlength{\tabcolsep}{4.5pt}
  \scriptsize
  \begin{tabular}{@{}L{2.0cm} C{1.65cm} | L{2.0cm} C{1.65cm}@{}}
    \toprule
    \rowcolor{hdrblue}
    \multicolumn{1}{c}{\textcolor{white}{\textbf{Parameter}}} &
    \multicolumn{1}{c|}{\textcolor{white}{\textbf{Value}}}    &
    \multicolumn{1}{c}{\textcolor{white}{\textbf{Parameter}}} &
    \multicolumn{1}{c}{\textcolor{white}{\textbf{Value}}}     \\
    \midrule
    \rowcolor{rowgray}
    \multicolumn{4}{l}{\textit{Channel \& Geometry}}                        \\
    Carrier freq.\ $f_c$     & $300\GHz$
      & Room dimensions         & $20{\times}20{\times}3\,\text{m}$          \\
    \rowcolor{rowgray}
    Bandwidth $B$            & $10\GHz$
      & Height $d$              & $3\,\text{m}$                               \\
    \multicolumn{4}{l}{\textit{Noise Model}}                                 \\
    \rowcolor{rowgray}
    Ref.\ temp.\ $T_0$       & $290\,\text{K}$
      & System temp.\ $\Tsys$  & ${\approx}1540\,\text{K}$                  \\
    Noise figure NF          & $8\dB$
      & Monte Carlo drops       & $2000$                                      \\
    \rowcolor{rowgray}
    \multicolumn{4}{l}{\textit{Molecular Absorption}}                        \\
    Shape $\varepsilon_K$    & $4$
      & Scale $\eta_K$         & $0.01\Npm$                                  \\
    \rowcolor{rowgray}
    Mean $\bar{K}$           & $0.04\Npm$
      & Std.\ dev.\ $\sigma_K$ & $0.02\Npm$                                 \\
    \multicolumn{4}{l}{\textit{THz-PASS Parameters}}                         \\
    \rowcolor{rowgray}
    Eff.\ index $\neff$    & $1.53$                                        
    & Atten.\ $\alpha$ & $0.05\,\text{dB/m}$    \\
       Waveguide length        & $20\,\text{m}$                                        
      &    Actuated PAs $\Nact$          & $8$                                  \\
    \rowcolor{rowgray}
    \multicolumn{4}{l}{\textit{Phased Array (UPA) Parameters}}               \\
    Element counts $M$       & $16/64/256$
      & Element spacing         & $\lambda_0/2$                               \\
    \rowcolor{rowgray}
    Element gain $G_t$       & $0\dBi$
      & UE gain $G_r$           & $5\dBi$                                     \\
    \multicolumn{4}{l}{\textit{Transmit Power \& Simulation}}                \\
    \rowcolor{rowgray}
    Power sweep $\Pt$        & $0$--$25\dBm$
      & Reference $\Pt$        & $10\dBm$                                     \\
    \bottomrule
  \end{tabular}
\end{table}

\begin{figure*}[!tbph]
  \centerline{
    \subfloat[]{%
      \includegraphics[width=0.45\textwidth]{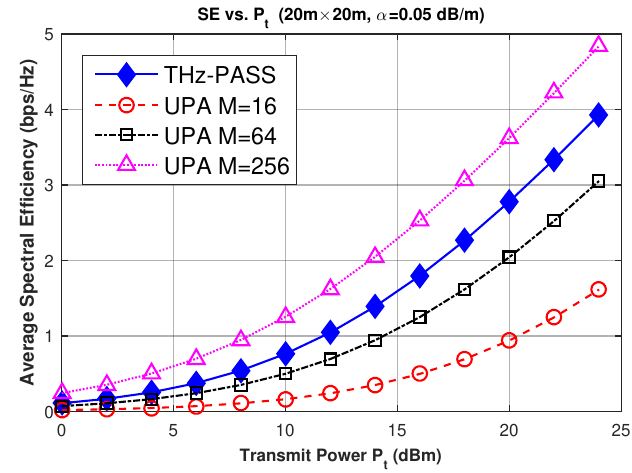}%
      \label{fig:se_pt}%
    }%
    \hspace{10mm}%
    \subfloat[]{%
      \includegraphics[width=0.45\textwidth]{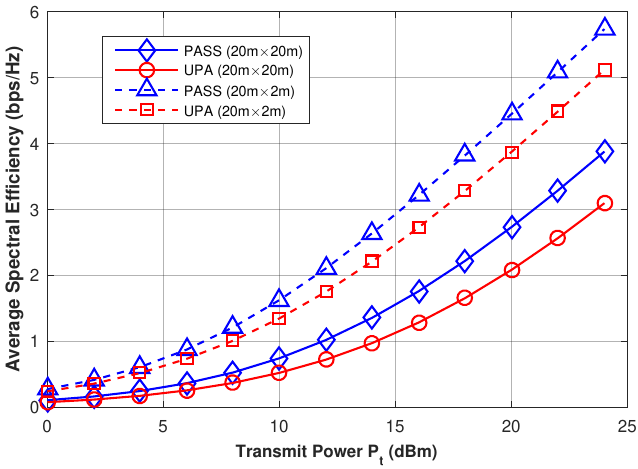}%
      \label{fig:scenario}%
    }%
  }
  \caption{Performance of THz-PASS compared against UPA\@:
    (a)~Spectral efficiency vs.\ transmit power $\Pt$
        ($20\times20\,\text{m}$, $\alpha=0.05\,\text{dB/m}$); and
    (b)~Average SE vs.\ $\Pt$ for open-office
        ($20\times20\,\text{m}$) and corridor ($20\times2\,\text{m}$)
        scenarios.}
  \label{Fig_performance}
\end{figure*}

Fig.~\ref{fig:se_pt} plots the average SE versus transmit power $\Pt$
for THz-PASS and three UPA configurations.
At $\Pt = 10\dBm$, THz-PASS achieves an average SE of
$0.746\,\text{bps/Hz}$, which outperforms UPA\,$M=16$
($0.163\,\text{bps/Hz}$, ${\times}4.6$ gain) and UPA\,$M=64$
($0.501\,\text{bps/Hz}$, $+1.73\dB$ gain), while remaining below
UPA\,$M=256$ ($1.252\,\text{bps/Hz}$).
The PASS advantage over UPA\,$M=64$ stems from the proximity effect:
a pinching antenna is always positioned directly above the served UE,
reducing the free-space path-loss distance from the array-to-UE
centroid distance (typically $\gg d$) to
$d_\mathrm{min} = \sqrt{\delta_\perp^2 + d^2}$,
where $\delta_\perp$ is the lateral UE offset from the waveguide.
In the $20\times20\,\text{m}$ room, $\delta_\perp$ is uniform on
$[0,\,D/2]$, yielding a mean $d_\mathrm{min}\approx 5.8\,\text{m}$
versus a UPA centroid distance of up to $14\,\text{m}$.
The SE curves grow logarithmically with $\Pt$ across the
$0$--$25\dBm$ range, with THz-PASS consistently surpassing
UPA\,$M=64$ beyond approximately $\Pt = 4\dBm$.
UPA\,$M=256$ dominates at high power because its 256-element
coherent array gain offsets the larger propagation distance, but
this comes at the cost of substantially higher hardware complexity,
power consumption, and circuit area.

To investigate the deployment-scenario dependence of THz-PASS,
Fig.~\ref{fig:scenario} compares two room geometries while fixing
$\alpha = 0.05\,\text{dB/m}$.
In the square room, UEs are distributed uniformly over a
$20\times 20\,\text{m}$ floor, resulting in a mean lateral distance to
the waveguide of $D/4 = 5\,\text{m}$.
THz-PASS achieves a mean SE of $0.746\,\text{bps/Hz}$ at $\Pt=10\dBm$.
Constraining the room width to $D = 2\,\text{m}$ represents a factory
aisle, aircraft cabin, or narrow office corridor, where the maximum
lateral UE distance from the waveguide is only $1\,\text{m}$.
In this case the PA-to-UE distance collapses to
$d_\mathrm{min}\approx\sqrt{1+9}=\sqrt{10}\approx 3.16\,\text{m}$ for
the worst-case UE and approaches $d=3\,\text{m}$ for all others.
The average SE more than doubles to $1.623\,\text{bps/Hz}$ for PASS
and to $1.353\,\text{bps/Hz}$ for UPA\,$M=64$.

Importantly, while UPA\,$M=64$ also benefits from the confined
geometry, the \emph{relative} PASS SE gain narrows from $+1.73\dB$
to $+0.79\dB$ in the corridor.
This occurs because UPA performance scales with the absolute path loss
(which improves in the corridor) independently of the waveguide
proximity advantage, partially closing the gap.
The corridor scenario therefore highlights a key deployment insight:
THz-PASS is \emph{particularly well-suited to confined or linear
topologies}—such as factory floors, transit corridors, aircraft
cabins, or server racks—where every UE lies close to the waveguide
axis and the proximity advantage is always fully exploited.

\begin{figure}[!tbph] 
    \centering
    \includegraphics[width=0.475\textwidth]{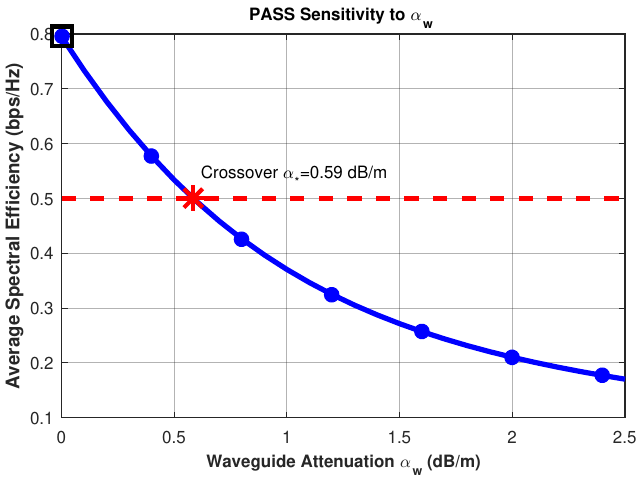}
     \caption{ THz-PASS average SE vs.\ waveguide attenuation $\alpha$ at $\Pt=10\dBm$; the crossover $\alpha^*$ marks the attenuation threshold above which UPA\,$M{=}64$ outperforms PASS.  }
     \label{fig:alpha_sens}    
\end{figure}

Fig.~\ref{fig:alpha_sens} illustrates how waveguide material quality
affects THz-PASS performance by sweeping from $\alpha=0.05\,\text{dB/m}$ to $\alpha=2.5\,\text{dB/m}$.
The PASS SE drops by approximately $30\%$ when the attenuation
coefficient degrades from $0.05$ to $0.5\,\text{dB/m}$.
This loss is attributable to the round-trip waveguide attenuation
$\exp(-2\alpha|x-x_\mathrm{feed}|)$: at $0.5\,\text{dB/m}$ and the
worst-case PA position ($|x - x_\mathrm{feed}| = 10\,\text{m}$), the
one-way power decays by $5\dB$, significantly attenuating the signal
delivered to the radiating pinching antenna.
The crossover threshold—where PASS SE equals UPA\,$M=64$—occurs at
$\alpha^* \approx 0.59\,\text{dB/m}$ under the open-office geometry,
as annotated in Fig.~\ref{fig:alpha_sens}.
Below this crossover, PASS outperforms UPA\,$M=64$ regardless of the
precise attenuation value, a condition that can be satisfied by
commercially available waveguide at 300\,GHz, see \cite{atakaramians2013terahertz}.

\section{CONCLUSIONs}

We have presented the first analytical framework for THz‑PASS, addressing the distinct electromagnetic phenomena that arise when extending pinching‑antenna systems to terahertz frequencies. Through extensive simulations at 300\GHz, we demonstrated that THz‑PASS outperforms 16‑ and 64‑element phased arrays under realistic indoor conditions, provided waveguide attenuation is below 0.59\dB/m. The fundamental advantage stems from proximity exploitation: pinching antennas positioned directly above users drastically reduce the effective propagation distance. Key design insights include the existence of an optimal number of elements due to mutual coupling, the critical role of waveguide material quality, and the particular suitability of THz‑PASS for confined linear topologies.


\begin{thebibliography}{99}
\bibitem{Ref_jiang2024terahertz}
W. Jiang, Q. Zhou, J. He, M. A. Habibi, S. Melnyk, M. El-Absi, B. Han, M. Di Renzo, H. D. Schotten, F.-L. Luo, T. S. El-Bawab, M. Juntti, M. Debbah, and V. C. M. Leung, ``Terahertz Communications and Sensing for {6G} and Beyond: A Comprehensive Review,'' in \emph{IEEE Commun. Surveys Tuts.}, vol. 26, no. 4, pp. 2326 - 2381, Fourthquarter 2024.

\bibitem{akyildiz2022terahertz}
I. F. Akyildiz, C. Han, Z. Hu, S. Nie, and J. M. Jornet, ``Terahertz Band Communication: An Old Problem Revisited and Research Directions for the Next Decade,'' \emph{IEEE Trans. Commun.}, vol. 70, no. 6, pp. 4250–4285, Jun. 2022.
\bibitem{molisch2017hybrid}
A. F. Molisch, V. V. Ratnam, S. Han, Z. Li, S. L. H. Nguyen, L. Li, and K. Haneda, ``Hybrid Beamforming for Massive MIMO: A Survey,'' \emph{IEEE Commun. Mag.}, vol. 55, no. 9, pp. 134–141, Sep. 2017.

\bibitem{jiang2023thz}
W. Jiang, and F.-L. Luo, ``Terahertz Technologies and Systems for 6G,'' in \emph{6G Key Technologies: A Comprehensive Guide}, 1st ed., Hoboken, NJ: Wiley-IEEE Press, 2023, ch. 5, pp. 195-252.

\bibitem{liu2025magazine}
Y. Liu, Z. Wang, X. Mu, C. Ouyang, X. Xu, and Z. Ding, ``Pinching-antenna systems: architecture designs, opportunities, and outlook,'' \emph{IEEE Commun. Mag.}, vol. 64, no. 1, pp. 190 - 196, Jan. 2026.

\bibitem{fukuda2021pinching}
A. Fukuda, H. Yamamoto, H. Okazaki, Y. Suzuki, and K. Kawai, ``Pinching Antenna — Using a Dielectric Waveguide as an Antenna,'' in \emph{NTT DOCOMO Technical J.}, vol. 23, no. 3, pp. 5 - 12, Jan. 2022.

\bibitem{ding2025flexible}
Z. Ding, R. Schober, and H. V. Poor, ``Flexible-Antenna Systems: A Pinching-Antenna Perspective,'' \emph{IEEE Trans. Commun.}, vol. 73, no. 10, pp. 9236 - 9253, Oct. 2025.

\bibitem{liu2025pinching}
Y. Liu, H. Jiang, X. Xu, Z. Wang, J. Guo, C. Ouyang, X. Mu,  Z. Ding, A. Nallanathan, G. K. Karagiannidis, and R. Schober, ``Pinching-Antenna Systems (PASS): A Tutorial,'' 
\emph{IEEE Trans. Commun.}, vol. 74, pp. 4881 - 4918, 2026.

\bibitem{ouyang2026multicast}
S. Shan, C. Ouyang, Y. Li, and Y. Liu, ``Exploiting Pinching-Antenna Systems in Multicast Communications,'' \emph{IEEE Trans. Commun.}, vol. 74, pp. 419 - 432, Oct. 2025.



\bibitem{wang2025modeling}
Z. Wang, C. Ouyang, X. Mu, Y. Liu, and Z. Ding, ``Modeling and Beamforming Optimization for Pinching-Antenna Systems,'' \emph{IEEE Trans. Commun.}, vol. 73, no. 12, pp. 13904–13919, Dec. 2025.
\bibitem{xu2025pinching}
X. Xu, X. Mu, Z. Wang, Y. Liu, and A. Nallanathan, ``Pinching-Antenna Systems (PASS): Power Radiation Model and Optimal Beamforming Design,'' \emph{IEEE Trans. Commun.}, vol. 74, pp. 2160–2175, Nov. 2025.



\bibitem{bhardwaj2024generalized}
P. Bhardwaj, R. Khanna, and S. M. Zafaruddin, ``A Generalized Statistical Model for THz Wireless Channel with Random Atmospheric Absorption,'' in \emph{Proc. IEEE Wireless Commun. Netw. Conf. (WCNC)}, Dubai, UAE, Apr. 2024, pp. 1–6.

\bibitem{jornet2011channel}
J. M. Jornet, and I. F. Akyildiz, ``Channel Modeling and Capacity Analysis for Electromagnetic Wireless Nanonetworks in the Terahertz Band,'' in \emph{IEEE Trans. Wireless Commun.}, vol. 10, no. 10, pp. 3211 - 3221, Oct. 2011.

\bibitem{ouyang2025ieee}
C. Ouyang, Z. Wang, Y. Liu, and Z. Ding, ``Array Gain for Pinching-Antenna Systems (PASS),'' \emph{IEEE Commun. Lett.}, vol. 29, no. 6, pp. 1471 - 1475, Jun. 2025. 

\bibitem{rappaport2019wireless}
T. S. Rappaport, Y. Xing, O. Kanhere, S. Ju, A. Madanayake, S. Mandal, A. Alkhateeb, and G. C. Trichopoulos, ``Wireless Communications and Applications Above 100 GHz: Opportunities and Challenges for 6G and Beyond,'' \emph{IEEE Access}, vol. 7, pp. 78729–78757, 2019.


\bibitem{vegni2016chirality}
A. M. Vegni and V. Loscrì, ``Chirality effects on channel modeling for THz-band wireless communications in LoS/NLoS propagation,'' \emph{Nano Commun. Netw.}, vol. 10, pp. 27–37, Dec. 2016.

\bibitem{atakaramians2013terahertz}
S. Atakaramians, S. Afshar, T. M. Monro, and D. Abbott, ``Terahertz dielectric waveguides,'' \emph{Adv. Opt. Photon.}, vol. 5, no. 2, pp. 169–215, Jul. 2013.

\end{thebibliography}
\end{document}